\documentclass[journal]{IEEEtran}
\usepackage{xspace,amsmath,amssymb,amsfonts,epsfig,subfigure,syntonly}
\usepackage{cite,bm,color,url,textcomp,empheq,boxedminipage}
\usepackage{algorithmicx,algorithm}
\usepackage{makecell}
\usepackage{empheq}
\usepackage{pifont}
\usepackage{soul}
\usepackage{graphicx,graphics}
\usepackage{multirow,multicol}
\usepackage{psfrag}
\usepackage{stfloats}
\usepackage{url}

\newcommand{\mv}[1]{\mbox{\boldmath{$ #1 $}}}
\newtheorem{remark}{\underline{Remark}}[section]

\long\def\symbolfootnote[#1]#2{\begingroup
\def\thefootnote{\fnsymbol{footnote}}
\footnote[#1]{#2}\endgroup}

\psfull

\begin{document}
\title{Cellular-Connected UAV with Adaptive Air-to-Ground Interference Cancellation and Trajectory Optimization}

\author{{Peiming Li,~\IEEEmembership{Graduate Student Member,~IEEE,} Lifeng Xie,~\IEEEmembership{Member,~IEEE,} Jianping Yao,~\IEEEmembership{Member,~IEEE,}\\ and Jie Xu,~\IEEEmembership{Member,~IEEE}}
\thanks{\scriptsize{P. Li and J. Yao are with the School of Information Engineering, Guangdong University of Technology, Guangzhou 510006, China (e-mail: peiminglee@outlook.com, yaojp@gdut.edu.cn). J. Yao is the corresponding author.}}
\thanks{\scriptsize{L. Xie is with the Peng Cheng Laboratory, Shenzhen 518000, China
(e-mail: xielf@pcl.ac.cn).}}
\thanks{\scriptsize{J. Xu is with the School of Science and Engineering and the Future Network of Intelligence Institute (FNii), The Chinese University of Hong Kong, Shenzhen, Shenzhen 518172, China (e-mail: xujie@cuhk.edu.cn).}}
}
\maketitle
\begin{abstract}
This letter studies a cellular-connected unmanned aerial vehicle (UAV) scenario, in which a UAV user communicates with ground base stations (GBSs) in cellular uplink by sharing the spectrum with ground users (GUs). To deal with the severe air-to-ground (A2G) co-channel interference, we consider an adaptive interference cancellation (IC) approach, in which each GBS can decode the GU's messages by adaptively switching between the modes of IC (i.e., precanceling the UAV's resultant interference) and treating interference as noise (TIN). By designing the GBSs' decoding mode, jointly with the wireless resource allocation and the UAV's trajectory control, we maximize the UAV's data-rate throughput over a finite mission period, while ensuring the minimum data-rate requirements at individual GUs. We propose an efficient algorithm to solve the throughput maximization problem by using the techniques of alternating optimization and successive convex approximation (SCA). Numerical results show that our proposed design significantly improves the UAV's throughput as compared to the benchmark schemes without the adaptive IC and/or trajectory optimization.
\end{abstract}
\section{Introduction}
Cellular-connected unmanned aerial vehicles (UAVs) have emerged as one of the key technologies for beyond fifth-generation (B5G) or sixth-generation (6G) cellular networks to enable secure and long-distance UAV applications \cite{JSAC,zeng1,zeng2}. UAVs can be connected with cellular networks as a new type of aerial users, which can share the scarce spectrum resources with conventional ground users (GUs). {As the UAVs fly at a relatively high altitude and normally have strong line-of-sight (LoS) links with ground base stations (GBSs) as compared to GUs, the resultant severe air-to-ground (A2G) co-channel interference from UAVs to GBSs is becoming a key technical challenge faced in cellular uplink integrated with UAVs \cite{R1}.}

In the literature, there have been various interference mitigation approaches proposed to tackle the A2G interference issue (see, e.g., \cite{weidong3}). {{For instance, the authors in \cite{weidong2} and \cite{R3} proposed an inter-cell interference coordination approach jointly designed with the UAV-GBS association.}} \cite{guiguan} investigated a local interference cancellation (IC) approach, in which the associated GBS can first decode the UAV's messages and precancel the resultant interference to facilitate the decoding of GUs' messages. Furthermore, \cite{weidong,weidong4,liuliang} investigated a cooperative IC method, in which GBSs can decode the UAV's messages and then send them to nearby GBSs to enable their cooperative IC, at the cost of data sharing over backhaul links. Despite their benefits, these prior works only considered one-shot resource allocation by considering UAVs staying at fixed locations, in which their controllable mobility was overlooked.

By exploiting the controllable mobility, the trajectory design has been recognized as a unique new design degree of freedom for optimizing the performance of UAV communications \cite{zeng2,JSAC}. {For UAV-enabled base stations (BSs), the authors in \cite{anti1} and \cite{anti2} considered that the UAV-BS is scheduled to collect data among ground nodes, and proposed different trajectory design schemes to achieve jamming resistance.
For cellular-connected UAVs, the authors in \cite{yuwei} and \cite{RL} considered the UAV trajectory optimization to maximize the UAV user's data-rate throughput over a certain mission period under different setups. However, only interference coordination with power control was considered, in which the A2G interference was treated as noise.}
\begin{figure}[t]
\centering
    \includegraphics[width=7cm]{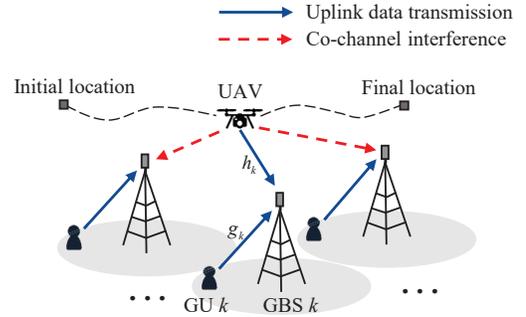}
\caption{{Illustration of the uplink spectrum sharing scenario for cellular-connected UAV.}} \label{model}
\end{figure}

To unlock the full potential of cellular-connected UAV, this letter exploits both benefits of IC and trajectory design to maximize the UAV communication performance. In particular, we consider an uplink spectrum sharing scenario, in which a UAV user communicates with GBSs by sharing the scarce spectrum resources with GUs. {We consider a new adaptive A2G IC approach, in which one or more GBSs are enabled to adaptively decode the UAV's messages based on the channel conditions, such that they can precancel the resultant A2G interference to facilitate the decoding of the GU's messages. Different from the cooperative IC method in \cite{weidong,weidong4,liuliang}, this approach only requires local IC at multiple GBSs, thus avoiding the data sharing cost over backhauls.{\footnote{In the isolated case, the data sharing among different GBSs is generally infeasible. If different GBSs are connected with fiber links, then they may be able to efficiently share their data to enable more advanced cooperative IC. This causes overhead in data exchange. In general, there exists a tradeoff between the interference cancellation gain and the overhead cost. However, as we do not consider the backhaul cost, how to evaluate it is an interesting topic that is left for future work.}} Under this setup, we consider a particular UAV mission period, and adaptively optimize each GBS's decoding mode between IC and treating interference as noise (TIN), jointly with the wireless resource allocation and the UAV's trajectory control. Our objective is to maximize the data-rate throughput of the UAV, while ensuring the minimum data-rate requirements at individual GUs. Although the formulated throughput maximization problem is highly non-convex, we propose an efficient algorithm by using the techniques of alternating optimization and successive convex approximation (SCA). In this algorithm, under any given UAV trajectory, we find the globally optimal resource allocation solution; while under given resource allocation, we obtain a converged trajectory solution. As such, the algorithm is ensured to converge. Numerical results show that our proposed joint design substantially improves the UAV's throughput as compared to the benchmark schemes without the adaptive IC and/or trajectory optimization.}
\section{System Model}\label{sec:II}
In this paper, we consider the uplink spectrum sharing scenario with one cellular-connected UAV user, where there are $K$ GBSs each serving one GU by using the same frequency band with the UAV, as shown in Fig. \ref{model}. Let $\mathcal{K}\triangleq\{1,\ldots,K\}$ denote the set of GBSs or their associated GUs. Suppose that each GBS $k\in{\mathcal K}$ locates at a fixed location $(x_k,y_k,0)$ on the ground in a three-dimensional (3D) coordinate system, where ${\mv \nu}_k= (x_k, y_k)\in \mathbb{R}^{2\times1}$ denotes the horizontal location.

We focus on a particular mission period ${\mathcal T}\triangleq[0,T]$, with finite duration $T$ in second (s), in which the UAV flies horizontally at a fixed altitude $H\geq0$. For ease of exposition, we discretize the whole period ${\mathcal T}$ into $N$ time slots each with a given duration $\delta_t=T/N$, where $\delta_t$ is sufficiently small such that the UAV's location is approximately unchanged during each slot.
Let $({x}[n],{y}[n],H)$ denote the time-varying location of the UAV at time slot $n\in{\mathcal N}\triangleq\{1,...,N\}$, where ${{\mv u}}[n]=({x}[n],{y}[n])$ denotes the UAV's horizontal location. Let ${\mv u}_\textrm{I}$ and ${\mv u}_\textrm{F}$ denote the UAV's horizontal initial and final locations, respectively, which are predetermined based on the specific UAV mission.
As a result, we have the following UAV flight constraints:
\begin{align}
&\|\mv u[n]-\mv u[n-1]\|\leq V_{\text{max}}\delta_t,\forall n\in{\mathcal N},\\
&{\mv u[0]}={\mv u}_\textrm{I},\ {\mv u}[N]={\mv u}_\textrm{F},
\end{align}
where $\|\cdot\|$ denotes the Euclidean norm, and $V_{\text{max}}$ denotes the maximum UAV speed.

We consider a quasi-static channel model, in which the wireless channels remain unchanged over each time slot. {In general, the A2G wireless links are mainly dominated by the LoS propagation due to the UAV's relatively high flight altitude \cite{channel}}. Therefore, we denote the channel power gain from the UAV to each GBS $k\in{\mathcal K}$ as
\begin{align}
h_{k}({\mv u}[n])\!=\!{\beta_{0}}d_k^{-\alpha}({\mv u}[n])\!=\!{\beta_{0}}/{(H^2\!+\!\|{\mv u}[n]\!-\!{\mv \nu}_k\|^2)^{\alpha/2}},
\end{align}
where $\alpha\ge2$ denotes the pathloss exponent, $\beta_0$ denotes the channel power gain at the reference distance of $d_0=1$ m, and $d_k({\mv u}[n])=\sqrt{H^2+\|{\mv u}[n]-{\mv \nu}_k\|^2}$ denotes the distance from the UAV to GBS $k\in{\mathcal K}$ at time slot $n\in{\mathcal N}$. Furthermore, we assume that the GUs stay at slow-varying or fixed locations. Accordingly, we denote ${g}_{k}$ as the channel power gain between GBS $k\in{\mathcal K}$ and its associated GU, which is assumed to remain unchanged over the mission period, and thus can be known by the GBSs {\it a-priori} to facilitate the adaptive IC and trajectory design. Furthermore, let ${q}_{k}[n]\ge0$ and $p[n]\ge0$ denote the transmit powers of GU $k$ and the UAV at time slot $n\in {\mathcal N}$, and $P$ and $Q_k$ denote their maximum transmit powers, respectively. Thus, we have $p[n]\le P$ and $q_k[n]\le Q_k,\forall k\in{\mathcal K},n\in{\mathcal N}$.

Next, we consider the cellular-connected UAV communication. Suppose that the UAV adopts the adaptive rate transmission, by setting the transmission rate as ${r}[n]\geq0$, which is a variable to be optimized later. Accordingly, at each time slot $n$, each GBS can determine whether to decode the UAV's messages or treat them as noise.
Let $\tau_{k}[n]\in\{0,1\}$ denote the decoding mode of GBS $k\in{\mathcal K}$ at time slot $n\in{\mathcal N}$, where $\tau_{k}[n]=1$ means that GBS $k$ needs to decode the UAV's messages (with IC) and $\tau_{k}[n]=0$ otherwise (with TIN). To ensure the UAV's communication, at least one GBS should successfully decode the UAV's messages at each time slot $n\in{\mathcal N}$, and thus we have $\sum\nolimits_{k\in{\mathcal K}}\tau_{k}[n]\ge1,\forall n\in{\mathcal N}$. Notice that the achievable rate from the UAV to GBS $k\in{\mathcal K}$ at time slot $n\in{\mathcal N}$ is given by
\begin{align}
{R}_k(p[n],{\mv u}[n],q_k[n])=\log_2\left(1+\frac{h_k({\mv u}[n])p[n]} {\sigma_k^2+{q}_{k}[n]{g}_{k}}\right),\label{Rkkkk}
\end{align}
where $\sigma_k^2$ denotes the background noise power at GBS $k$ containing the potential terrestrial inter-cell interference from other GUs. {\footnote{{ Note that in practice the ground-to-ground (G2G) inter-cell interference depends on various issues such as the location of the GUs in the nearby cells and the transmit power of that user. Here, the background noise power $\sigma^2_k$ is considered to be constant to facilitate the offline trajectory design to characterize the performance limits, which can be appropriately set and served as an upper bound of the potential G2G inter-cell interference in practical implementation. Accordingly, these GUs need to implement the online transmit power control to meet such interference limitations, which is left for future research.}}}
Therefore, it must follow that
\begin{align}
\tau_{k}[n]r[n]\leq R_k(p[n],{\mv u}[n],q_k[n]),\forall k\in{\mathcal K},n\in{\mathcal N}.\label{Rk1}
\end{align}

Then, we consider the decoding of the associated GU $k$'s messages at each GBS $k\in {\mathcal K}$. Consider a particular time slot $n\in{\mathcal N}$. If $\tau_k[n]= 1$ holds with the UAV's messages successfully decoded, then GBS $k$ can adopt IC, i.e., GBS $k$ can first cancel the UAV's resulting interference, and then decode the GU's messages. In this case, the achievable rate of GU $k$ at time slot $n$ is
\begin{align}
\bar{R}_k^{\text{IC}}(q_k[n])=\log_2(1+g_kq_k[n]/\sigma^2_k).\label{RIC}
\end{align}
On the other hand, if $\tau_k[n]= 0$, then GBS $k$ needs to decode the messages of GU $k$ by TIN (i.e., treating the interference from the UAV as noise). As a result, its achievable rate is given by
\begin{align}
\bar{R}_k^{\text{TIN}}(p[n],{\mv u}[n],q_k[n])\!=\!\log_2\left(1\!+\!\frac{g_kq_k[n]}{\sigma^2_k\!+\!h_k({\mv u}[n])p[n]}\right).\label{RTIN}
\end{align}

Our objective is to maximize the UAV's average throughput (i.e., $\frac{1}{N}\sum_{n\in{\mathcal N}}r[n]$) over the mission period, while ensuring the GUs' communication requirements. In particular, we suppose that at each time slot $n\in {\mathcal N}$, the achievable rate of each GU must be no less than a certain threshold $\Gamma_k \ge 0$. The decision variables include the UAV trajectory $\{{\mv u}[n]\}$, the transmit power of the UAV $\{p[n]\}$ and GUs $\{q_k[n]\}$, the decoding mode of GBSs $\{\tau_{k}[n]\}$, as well as the UAV's {transmission rate $\{r[n]\}$}. Therefore, the throughput maximization problem of our interest is formulated as
\begin{align}
\text{(P1):}&\max_{\substack{\{\tau_{k}[n]\},\{q_k[n]\},\{p[n]\},\{{\mv u}[n]\},\{r[n]\ge0\}}}~\frac{1}{N}\sum\nolimits_{n\in{\mathcal N}}r[n]\notag\\
\text{s.t.}~&\|\mv u[n]-\mv u[n-1]\|\leq V_{\text{max}}\delta_t,\forall n\in{\mathcal N}\label{eqn:speed}\\
&{\mv u[0]}={\mv u}_\textrm{I},\ {\mv u}[N]={\mv u}_\textrm{F}\label{eqn:inifin}\\
&0\le p[n]\le P,~0\le q_k[n]\le Q_k,\forall k\in{\mathcal K},n\in{\mathcal N}\label{power}\\
&\sum\nolimits_{k\in{\mathcal K}}\tau_k[n]\ge1,~\tau_{k}[n]\in\{0,1\},\forall k\in{\mathcal K},n\in{\mathcal N}\label{tau}\\
&\tau_{k}[n]{r}[n]\le{R}_{k}(p[n],{\mv u}[n],q_k[n]), \forall k\in{\mathcal K},n\in{\mathcal N}\label{rate}\\
&\tau_k[n]\bar{R}_k^{\text{IC}}(q_k[n])+(1-\tau_k[n])\bar{R}_k^{\text{TIN}}(p[n],{\mv u}[n],q_k[n])\notag\\
&~~~~~~~~~~~~~~~~~~~~~~~~~~~~~~~\geq\Gamma_k,\forall k\in{\mathcal K},n\in{\mathcal N}.\label{IT}
\end{align}
Here, \eqref{eqn:speed} and \eqref{eqn:inifin} denote the UAV flight constraints, \eqref{power} denotes the power constraints for both the UAV and GUs, \eqref{tau} denotes the decoding mode constraints for GBSs, \eqref{rate} denotes the rate constraints for the UAV, and \eqref{IT} denotes the minimum rate constraints for GUs. Note that problem (P1) is a mixed-integer non-convex problem, as the variable $\{\tau_k[n]\}$ are binary, and the constraints in \eqref{rate} and \eqref{IT} are non-convex due to the coupling of variables.

Before proceeding, we check the feasibility of problem (P1). {Notice that problem (P1) is feasible, as long as the mission duration $T$ is sufficient for the UAV to fly from the initial location to the final location (i.e., $\|{\mv u}_\text{F} - {\mv u}_\text{I}\| \leq V_\text{max}T$) and the communication requirements of GUs can be met at their maximum powers under the IC mode (i.e., $\bar{R}_k^{\text{IC}}(Q_k)\geq\Gamma_k,\forall k\in{\mathcal K}$).{{\footnote{{It is observed from \eqref{RIC} and \eqref{RTIN} that the IC mode is able to achieve no smaller data rate than that by TIN mode, provided that the rate constraints in \eqref{Rk1} are met. Notice that we can always make \eqref{Rk1} satisfied by setting the UAV's transmission rate $r[n]$ to be zero. Therefore, it is sufficient to consider the IC mode for checking the feasibility of problem (P1).}}}} In practice, the UAV mission duration $T$ cannot exceed the UAV's maximum battery lifetime $T_\text{max}$. Supposing that $T$ and $\{Q_k\}$ are sufficiently large, we focus on the case when problem (P1) is feasible in the sequel.}
\section{Proposed Solution to Problem (P1)}\label{sec:solution}
In this section, we propose an efficient algorithm to solve problem (P1) iteratively by using the technique of alternating optimization. In particular, in each iteration we optimize the resource allocation (i.e., the decoding mode of GBSs $\{\tau_k[n]\}$, and the power control of GUs $\{q_k[n]\}$ and the UAV $\{p[n]\}$), as well as the UAV trajectory $\{{\mv u}[n]\}$, in an alternating manner, by assuming the other to be given.
\subsection{Resource Allocation Optimization for (P1) Under Given UAV Trajectory}
Under any given UAV trajectory $\{{\mv u}[n]\}$, problem (P1) is reduced to
\begin{align}
&\text{(P1.1):}\max_{\substack{\{\tau_{k}[n]\},\{q_k[n]\},\{p[n]\},\{r[n]\ge0\}}}~\frac{1}{N}\sum\nolimits_{n\in{\mathcal N}}r[n]\notag\\
&~~~~~~~~~~~~~~~~~~~~~~~\text{s.t.}~\eqref{power},\eqref{tau},\eqref{rate},\eqref{IT}.\notag
\end{align}
It is observed that problem (P1.1) can be decomposed into $N$ subproblems in problem (P1.2), each corresponding to optimizing $\{\tau_k[n]\}$, $\{q_k[n]\}$, $p[n]$, and $r[n]$ at time slot $n\in{\mathcal N}$ with a given UAV location ${\mv u}[n]$.
\begin{align}
&\text{(P1.2):}\max_{\substack{\{\tau_{k}[n]\},\{q_k[n]\},p[n],r[n]\ge0}}~r[n]\notag\\
&\text{s.t.}~0\le p[n]\le P,~0\le q_k[n]\le Q_k,\forall k\in{\mathcal K}\label{power11}\\
&~~~~\sum\nolimits_{k\in{\mathcal K}}\tau_k[n]\ge1,~\tau_{k}[n]\in\{0,1\},\forall k\in{\mathcal K}\label{tau11}\\
&~~~~\tau_{k}[n]{r}[n]\le{R}_{k}(p[n],{\mv u}[n],q_k[n]), \forall k\in{\mathcal K}\label{rate11}\\
&~~~~\tau_k[n]\bar{R}_k^{\text{IC}}(q_k[n])+(1-\tau_k[n])\bar{R}_k^{\text{TIN}}(p[n],{\mv u}[n],q_k[n])\notag\\
&~~~~~~~~~~~~~~~~~~~~~~~~~~~~~~~~~~~~~~~~~~~~~\geq\Gamma_k,\forall k\in{\mathcal K}.\label{IT11}
\end{align}
To solve problem (P1.2), we first solve for $\{q_k[n]\}$, $p[n]$, and $r[n]$ by considering the decoding mode $\{\tau_k[n]\}$ to be fixed, and then compare the achieved objective values under different $\{\tau_k[n]\}$ to obtain the optimal solution.

First, consider the case when $\{\tau_k[n]\}$ are given. In this case, let ${\mathcal K}^\text{IC} = \{k \in {\mathcal K}| \tau_k[n] = 1 \}$ denote the set of GBSs adopting the IC mode and ${\mathcal K}^\text{TIN} = \{k \in {\mathcal K}| \tau_k[n] = 0 \}$ denote that adopting the TIN mode. If GBS $k\in{\mathcal K}^\text{IC}$, based on constraints \eqref{rate11} and \eqref{IT11}, the optimal power allocation of GU $k\in{\mathcal K}^\text{IC}$ is
\begin{align}
q^\text{*}_k[n] = (2^{\Gamma_k}-1)\sigma^2_k/g_k, \forall k\in {\mathcal K}^\text{IC}.\label{powerIC}
\end{align}
In this case, constraints \eqref{rate11} and \eqref{IT11} respectively become
\begin{align}
&{r}[n]\leq{R}_{k}(p[n],{\mv u}[n],q_k^\text{*}[n]),\forall k\in{\mathcal K}^\text{IC},\label{AA}\\
&{\bar R}_k^\text{TIN}(p[n],{\mv u}[n],q_k[n])\geq \Gamma_k, \forall k\in {\mathcal K}^\text{TIN}.\label{BB}
\end{align}

As a result, problem (P1.2) under given $\{\tau_k[n]\}$ is reduced as
\begin{align}
&\text{(P1.3):}\max_{{\{q_k[n]\}, p[n],r[n]\ge0}}~r[n]\notag\\
&~~~~~~~~~~~~~~~~\text{s.t.}~\eqref{power11},\eqref{AA},\eqref{BB}.\notag
\end{align}
It is easy to observe that at the optimality of problem (P1.3), the constraints in \eqref{BB} must be met with strict equality, i.e.,
\begin{align}
\bar{R}_{{k}}^{\text{TIN}}(p[n],{\mv u}[n],q_{{k}}[n])=\Gamma_{{k}},\forall k\in{\mathcal K},\label{R_k}
\end{align}
and accordingly, we have
\begin{align}
p[n] = \frac{g_{k}q_{k}[n]}{(2^{\Gamma_{k}}-1)h_{{k}}({\mv u}[n])}-\frac{\sigma^2_{k}}{h_{{k}}({\mv u}[n])},\forall k\in{\mathcal K}\label{DD}.
\end{align}
Furthermore, by combining \eqref{DD} with constraint \eqref{power11}, we have the optimal power allocation of GU $k\in{\mathcal K}^\text{TIN}$ as
\begin{align}
q^\text{*}_k[n] = Q_k, \forall k\in {\mathcal K}^\text{TIN},\label{powerTIN}
\end{align}
and
\begin{align}
p[n] \le \min\nolimits_{k\in {\mathcal K}^\text{TIN}}\left(\frac{g_{k}Q_{k}}{(2^{\Gamma_{k}}-1)h_{{k}}({\mv u}[n])}-\frac{\sigma^2_{k}}{h_{{k}}({\mv u}[n])}\right).\label{powerA}
\end{align}
Notice that the achievable rate of the UAV or the objective value of problem (P1.3) is monotonically increasing with respect to $p[n]$. Therefore, based on \eqref{powerA} and  $p[n]\le P$, we obtain the optimal solution of $p[n]$ as
\begin{align}
&p^*[n]\!=\!\min\!\left(\min\limits_{k\in{\mathcal K}^\text{TIN}}\!\left(\frac{g_{k}Q_{k}}{(2^{\Gamma_{k}}\!-\!1)h_{{k}}({\mv u}[n])}\!-\!\frac{\sigma^2_{k}}{h_{{k}}({\mv u}[n])}\right),P\right).\label{power*}
\end{align}
Accordingly, based on \eqref{AA}, the optimal rate $r[n]$ of the UAV is given as
\begin{align}
r^*[n]=\min\nolimits_{k\in{\mathcal K}^\text{IC}}\left({R}_{k}(p^*[n],{\mv u}[n],q_k^\text{*}[n])\right).\label{opt-rate}
\end{align}
By combining \eqref{powerIC}, \eqref{powerTIN}, \eqref{power*}, and \eqref{opt-rate}, the optimal solution to problem (P1.2) with fixed $\{\tau_k[n]\}$ is obtained.

Next, we compare the obtained achievable rate or optimal objective value in \eqref{opt-rate} of problem (P1.3) under different values of $\{\tau_k[n]\}$. Notice that due to the constraints in \eqref{tau11}, there are a total number of $2^K-1$ options with at least one of the $\tau_k[n]$'s must be one. By comparing the $2^K-1$ optimal values, we can get the optimal decoding mode solution to problem (P1.2) as $\{\tau^\star_k[n]\}$.
Accordingly, the optimal solution $\{q_k^*[n]\}$, $p^*[n]$, and $r^*[n]$ in \eqref{powerIC}, \eqref{powerTIN}, \eqref{power*}, and \eqref{opt-rate} under the optimal $\{\tau^\star_k[n]\}$ correspond to the optimal solution $\{q_k^\star[n]\}$, $p^\star[n]$, and $r^\star[n]$ to problem (P1.2). As such, problem (P1.2) is optimally solved. Accordingly, problem (P1.1) is solved.
\subsection{UAV Trajectory Optimization for (P1) Under Given Resource Allocation}
Under any given resource allocation $\{\tau_k[n]\}$, $\{q_k[n]\}$, and $\{p[n]\}$, problem (P1) is reduced as the following trajectory optimization problem:
\begin{align}
&\text{(P1.4):}\max_{\substack{\{{\mv u}[n]\},\{r[n]\ge0\}}}~\frac{1}{N}\sum\nolimits_{n\in{\mathcal N}}r[n]\notag\\
&~~~~~~~~~~~~~~~\text{s.t.}~\eqref{eqn:speed},\eqref{eqn:inifin},\eqref{rate},\eqref{IT}.\notag
\end{align}
Notice that problem (P1.4) is still a non-convex optimization problem, as the constraints in \eqref{rate} and \eqref{IT} are non-convex with respect to $\{{\mv u}[n]\}$. Therefore, problem (P1.4) cannot be solved by standard convex optimization techniques. To tackle this difficulty, we adopt the SCA-based algorithm to obtain an efficient solution, which is implemented in an iterative manner as follows.

Consider each iteration $j \ge 1$, in which the local trajectory point is denoted as $\{{\mv u}^{(j)}[n]\}$. Accordingly, the non-convex constrains in \eqref{rate} and \eqref{IT} can be approximated into convex forms as follows. First, we consider constraint \eqref{rate}. Note that ${R}_{k}(p[n],{\mv u}[n],q_k[n])$ in \eqref{Rkkkk} is a convex function with respect to $\|{\mv u}[n]-{\mv \nu}_k\|^2$. As the first-order Taylor expansion of a convex function is a global underestimate of the function value, we have
\begin{align}
{R}_{k}(p[n],{\mv u}[n],q_k[n])\ge \hat{R}_{k}^{\text{lb}(j)}(p[n],{\mv u}[n],q_k[n]),
\end{align}
with
\begin{align}
&\hat{R}_{k}^{\text{lb}(j)}(p[n],{\mv u}[n],q_k[n])\triangleq{R}_{k}(p[n],{\mv u}^{(j)}[n],q_k[n])\notag\\
&~~~~~~~~~-A_{k}^{(j)}[n](\|{\mv u}[n]-{\mv \nu}_k\|^2-\|{\mv u}^{(j)}[n]-{\mv \nu}_k\|^2),
\end{align}
where $A_{k}^{(j)}[n]=\frac{1}{2}\alpha p[n]\beta_0/\ln2(p[n]\beta_0d_{k}({\mv u}^{(j)}[n])+(\sigma^2_{k}+g_{k}q_{k}[n])d^{{\alpha}+1}_{k}({\mv u}^{(j)}[n]))$.
Therefore, the non-convex constraints in \eqref{rate} can be approximated as the following convex constraints:
\begin{align}
\tau_{k}[n]{r}[n]\le \hat{R}_{k}^{\text{lb}(j)}(p[n],{\mv u}[n],q_k[n]), \forall k\in{\mathcal K}, n\in{\mathcal{N}}\label{eqn:SCP1}.
\end{align}
Next, for~constraint~\eqref{IT},~we~rewrite~$\bar{R}_k^{\text{TIN}}(p[n],{\mv u}[n],q_k[n])$~as
\begin{align}
\bar{R}_k^{\text{TIN}}(p[n],{\mv u}[n],q_k[n])=&\log_2\left(\sigma^2_k+p[n]h_k({\mv u}[n])+q_k[n]g_k\right)\notag\\
&-\log_2\left(\sigma^2_k+h_k({\mv u}[n])p[n]\right).\label{expand}
\end{align}
Note that the first term at the right-hand-side (RHS) in \eqref{expand} is a convex functions with respect to $\|{\mv u}[n]-{\mv \nu}_k\|^2$. Similarly, we have
\begin{align}
&\log_2\left(\sigma^2_k+h_k({\mv u}[n])p[n]+g_kq_k[n]\right)\notag\\
&\geq\log_2\left(\sigma^2_k+h_k({\mv u}^{(j)}[n])p[n]+g_kq_k[n]\right)\notag\\
&~~~-B^{(j)}_k[n](\|{\mv u}[n]-{\mv \nu}_k\|^2-\|{\mv u}^{(j)}[n]-{\mv \nu}_k\|^2)\notag\\
&\triangleq \check{R}_k^{\text{lb}(j)}(p[n],{\mv u}[n],q_k[n]),
\end{align}
where $B_{k}^{(j)}[n]=\frac{1}{2}\alpha p[n]\beta_0/\ln2((\sigma^2_k+p[n]\beta_0d^{-\alpha}_{k}({\mv u}^{(j)}[n])+q_{k}[n]g_{k})d^{{\alpha}+1}_{k}({\mv u}^{(j)}[n]))$.
Therefore, the non-convex constraints in \eqref{IT} can be approximated as the following convex constraints:
\begin{align}
&\tau_k[n]\bar{R}_k^{\text{IC}}(q_k[n])+(1-\tau_k[n])(\check{R}_k^{\text{lb}(j)}(p[n],{\mv u}[n],q_k[n])\notag\\
&~~~-\log_2(\sigma^2_k+h_k({\mv u}[n])p[n]))\geq\Gamma_k, \forall k\in{\mathcal K},n\in{\mathcal N}.\label{eqn:SCP2}
\end{align}

As a result, with given local point $\{{\mv u}^{(j)}[n]\}$, problem (P1.4) is approximated as the following convex optimization problem (P1.5), which can thus be solved optimally via standard convex optimization techniques such as CVX \cite{CVX}.
\begin{align}
&\text{(P1.5):}\max_{\substack{\{{\mv u}[n]\},\{r[n]\ge0\}}}~\frac{1}{N}\sum\nolimits_{n\in{\mathcal N}}r[n]\notag\\
&~~~~~~~~~~~~~~~\text{s.t.}~\notag\eqref{eqn:speed},\eqref{eqn:inifin},\eqref{eqn:SCP1},\eqref{eqn:SCP2}.
\end{align}

The obtained optimal solution to problem (P1.5) under given local point $\{{\mv u}^{(j-1)}[n]\}$ is given as $\{{\mv u}^{(j)}[n]\}$, which is then used as the local point for the next iteration $j+1$. As the obtained objective value of problem (P1.4) is monotonically non-decreasing for each iteration and the optimal value is upper bounded, it is clear that the SCA-based update will lead to a converged solution to problem (P1.4).
\subsection{Complete Algorithm for Solving (P1)}
By combining the solutions in Sections \ref{sec:solution}-A and \ref{sec:solution}-B, we solve problem (P1) by updating the variables iteratively in an alternating manner. In each iteration, we first solve problem (P1.1) under given $\{{\mv u}[n]\}$ to derive the closed-form solutions of $\{\tau_{k}[n]\}$, $\{q_k[n]\}$, $\{p[n]\}$, and $\{{r}[n]\}$, and then solve problem (P1.4) based on SCA under the obtained $\{\tau_k[n]\}$, $\{q_k[n]\}$, and $\{p[n]\}$ to update $\{{\mv u}[n]\}$ and $\{r[n]\}$. {The complete algorithm for solving problem (P1) is summarized in Table \ref{t1}.
For each iteration, the updated objective value of problem (P1) is ensured to be monotonically non-decreasing, and as a result, the convergence of the proposed algorithm for problem (P1) is ensured. Please refer to Appendix \ref{Appendix} for the detailed convergence proof.}

\begin{table}[ht]
{
\begin{center}
\caption{{Complete Algorithm for Solving Problem (P1)}}\label{t1}
\hrule \vspace{1pt}
\begin{itemize}
\item[1.] {\bf Initialization}: Set the initial UAV trajectory as $\{\hat{\mv u}^{\left(0\right)}[n]\}$, and set $i=1$.
\item[2.] {\bf Repeat:}
\begin{itemize}
\item[1)] {\bf Solve problem (P1.1) for resource allocation optimization under given UAV trajectory $\{\hat{\mv u}^{\left(i-1\right)}[n]\}$}:
    \begin{itemize}
    \item[a.] Decompose problem (P1.1) into $N$ subproblems as problem (P1.2) for each time slot $n\in{\mathcal N}$;
    \item[b.] Solve problem (P1.2) under given decoding mode (i.e., problem (P1.3)), and obtain the optimal power allocation solution based on \eqref{powerIC}, \eqref{powerTIN}, \eqref{power*}, and \eqref{opt-rate};
    \item[c.] Compare the optimal objective value of problem (P1.3) under the $2^K-1$ decoding modes, and obtain the optimal decoding mode solution $\{\hat{\tau}_k^{(i)}[n]\}$ to problem (P1.2), then calculate the optimal solution of $\{q_k^\star[n]\}$, $p^\star[n]$, and $r^\star[n]$ under the optimal decoding mode $\{\hat{\tau}_k^{(i)}[n]\}$ to problem (P1.2);
    \item[d.] Obtain the resource allocation $\{\hat{q}_k^{(i)}[n]\}$ and $\hat{p}^{(i)}[n]$ during ${\mathcal T}$ based on the optimal solution to problem (P1.2) for each time slot $n\in{\mathcal N}$.
   \end{itemize}
\item[2)] {\bf Solve problem (P1.4) for UAV trajectory optimization under given resource allocation $\{\hat{\tau}^{\left(i\right)}_k[n]\}$, $\{\hat{q}^{\left(i\right)}_k[n]\}$, and $\{\hat{p}^{\left(i\right)}[n]\}$}, set $j=1$ and $\{{\mv u}^{(0)}[n]\}=\{\hat{\mv u}^{(i-1)}[n]\}$.
    \begin{itemize}
    \item[a.] {\bf Repeat:}
    \begin{itemize}
    \item[a)] Solve the convex optimization problem (P1.5) under given local point $\{{\mv u}^{(j-1)}[n]\}$, and obtain the optimal solution $\{{\mv u}^{\left(j\right)}[n]\}$ and $\{r^{\left(j\right)}[n]\}$;
    \item[b)]  $j=j+1$.
    \end{itemize}
    \item[b.] {\bf Until} converge or reach a maximum number of iterations, and set $\{\hat{\mv u}^{(i)}[n] \}= \{{\mv u}^{(j)}[n]\}$ and $\{\hat{r}^{(i)}[n] \}= \{r^{(j)}[n]\}$.
   \end{itemize}
\item[3)] $i=i+1$.
\end{itemize}
\item[3.] {\bf Until} converge or reach a maximum number of iterations.
\end{itemize}\vspace{1pt}
\hrule
\end{center}}
\end{table}

{{{\it Complexity Analysis:} First, the complexity of wireless resource allocation via solving (P1.2) is given by ${\mathcal O}(NK2^K)$. This is due to the fact that in each of the $N$ implementations, $\{\tau_k[n]\}$ are determined by comparing the obtained $2^K - 1$ options, where \eqref{power*} and \eqref{opt-rate} respectively have a complexity of ${\mathcal O}(K)$. Next, solving problem (P1.4) by solving its approximating problem (P1.5) via CVX \cite{boyd}, the interior point method has a complexity of ${\mathcal O}(K^{1.5}N^{3.5})$. By letting $L_\text{AO}$ and $L_\text{SCA}$ denote the numbers of iterations in the alternating optimization for problem (P1) and the SCA for problem (P1.4), respectively, the computation complexity of the proposed alternating optimization-based approach for solving problem (P1) is ${\mathcal O}(L_\text{AO}(NK2^K+L_\text{SCA}K^{1.5}N^{3.5}))$.}}

{\begin{remark}
The joint design of UAV trajectory and wireless resource allocation requires the coordination among the GBSs, GUs, and the UAV. This can be controlled by a centralized network controller (e.g., the cloud in the cloud radio access network (CRAN)). In practice, the centralized controller needs to collect both flight-related and channel state information of the wireless network with UAV integrations, and then implement the proposed algorithm to obtain the resource allocation and UAV trajectory. After that, the controller can send the corresponding information signals to all GBSs, GUs, and the UAV, which can then accordingly determine their transmission schemes to implement the cooperative design. Notice that in general such a scheduling process should be implemented in an offline manner to provide rough trajectory planning, which can be refined later based on an additional online design relying on real-time information \cite{online}. Such offline trajectory design methods have been commonly adopted in the literature, such as, e.g.,\cite{online1,online2,online3}. It is also worth noting that the considered interference cancellation design does not require data sharing among GBSs \cite{weidong,weidong4,liuliang}, and thus can be implemented independently on each GBS, without modifying the communication protocol. This is more effective in practical implementation.
\end{remark}}

%

\section{Numerical Results}\label{sec:V}
This section provides numerical results to validate the performance of the proposed joint design, as compared with the following five benchmark schemes.
\begin{itemize}
{\item{\it Upper bound with $T\rightarrow\infty$}: When $T\rightarrow\infty$, the UAV's maximum flying speed constraints in \eqref{eqn:speed} and the initial and final locations constraints in \eqref{eqn:inifin} can be ignored, and the resultant throughput becomes the performance upper bound. In this case, the corresponding optimization problem can be solved via a two-dimensional (2D) exhaustive search to find the optimal UAV hovering location together with the solution to problem (P1.1) to find the optimal resource allocation.{\footnote{{Notice that we cannot obtain the optimal solution under the general case of $T$, but we can obtain an upper bound by considering $T$ goes to infinity, as adopted in \cite{TSP}.}}}}
\item {\it Straight-fly scheme}: The UAV flies straight from the initial to the final locations at a uniform speed $\|{\mv u}_\text{F}-{\mv u}_\text{I}\|/T$. This scheme corresponds to solving problem (P1.1) under a given UAV trajectory. This is also adopted as the initial UAV trajectory for the proposed design.
{\item {\it Successive-hover-fly scheme}: In this scheme, the UAV trajectory is obtained by solving a traveling salesman problem (TSP) among the initial and final locations as well as three GBSs to find the shortest path to visit them. In addition, the hovering time at each GBS can be obtained by solving a linear program \cite{TSP}.}
\item {\it Egoistic decoding scheme}: In this scheme, only one GBS decodes the UAV's messages at each time slot $n\in{\mathcal N}$. This scheme corresponds to solving problem (P1) by replacing constraint \eqref{tau} as $\sum\nolimits_{k\in{\mathcal K}}\tau_k[n]=1,\forall n\in{\mathcal N}$.
\item {\it Altruistic decoding scheme}: In this scheme, all {GBSs} are enabled to decode the UAV's messages. This scheme corresponds to solving problem (P1) under fixed decoding mode with $\tau_k[n]=1,\forall k\in{\mathcal N}, n\in{\mathcal N}$, where constraint \eqref{rate} becomes
    ${r}[n]\le\min_{k\in{\mathcal K}}{R}_{k}(p[n],{\mv u}[n],q_k[n]), \forall n\in{\mathcal N}$.
\end{itemize}

\begin{figure}[t]
\centering
\includegraphics[width=6.5cm]{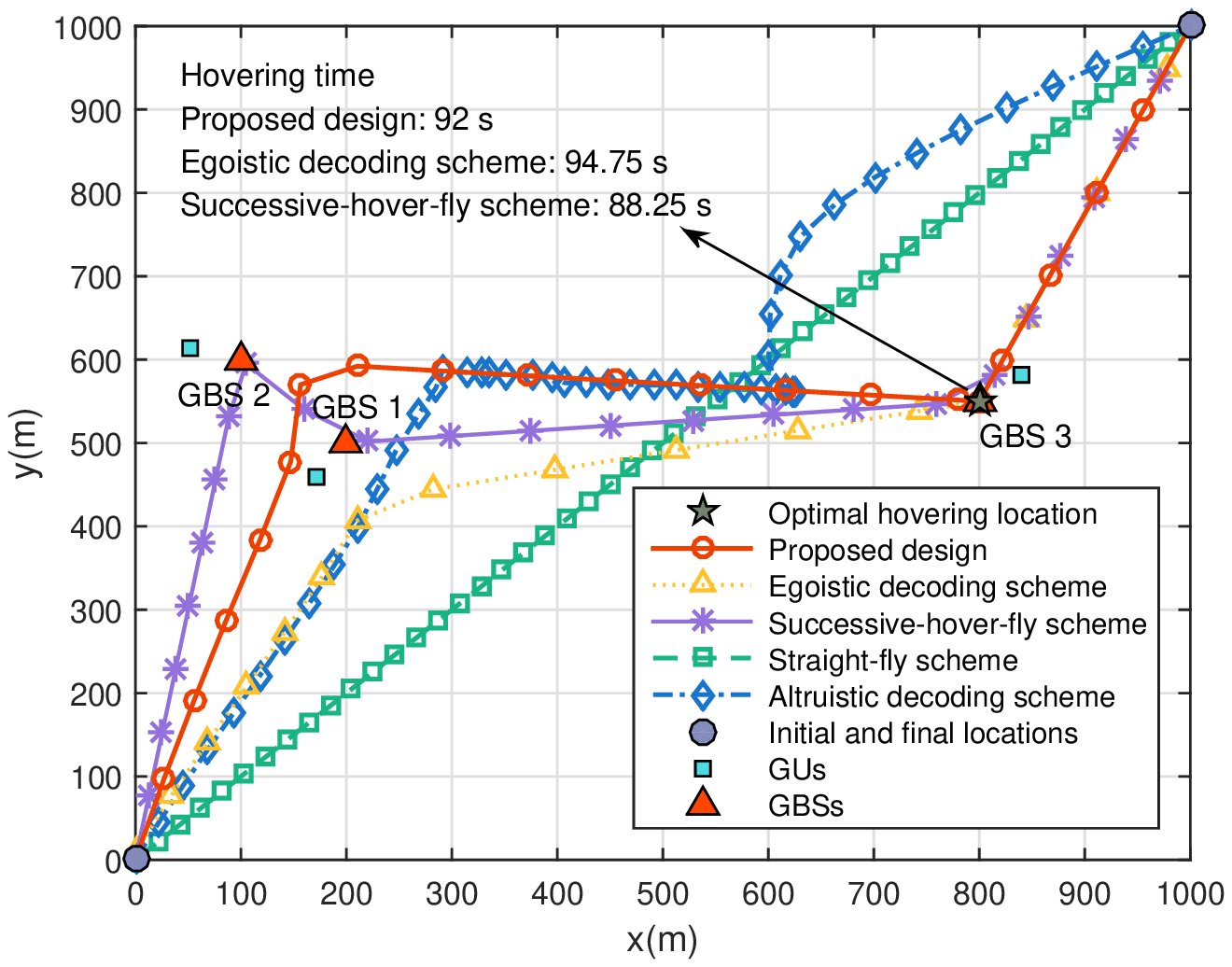}
\caption{{The optimized UAV trajectory projected on the horizontal plane with $T=150$ s.}}\label{2usertrajectory}
\includegraphics[width=6.5cm]{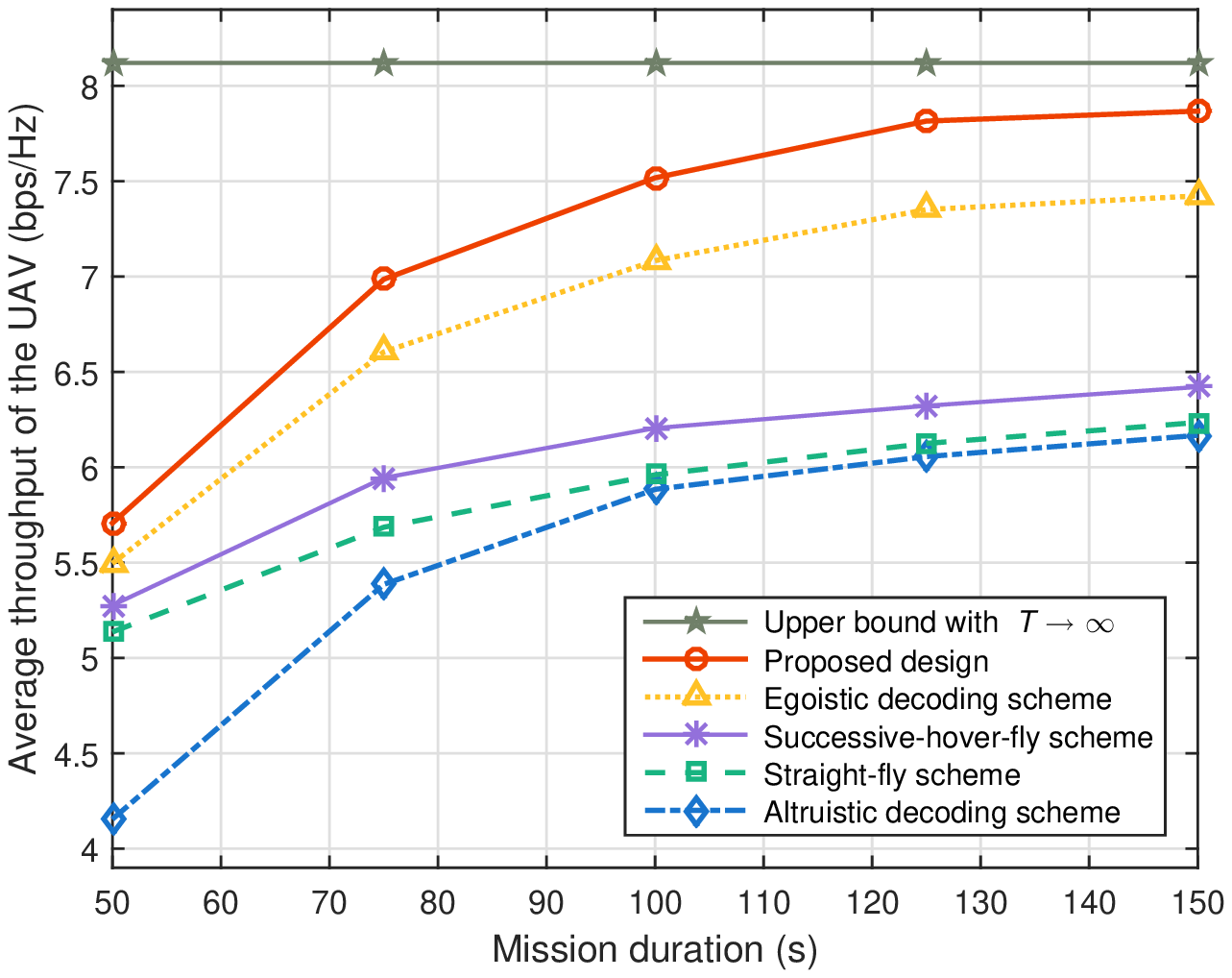}
\caption{{The average throughput of the UAV versus mission duration $T$.}}\label{2userperformance}
\end{figure}

In the simulation, suppose that there are $K=3$ {GBSs} and GUs distributed within a geographic area of size $1\times1~\text{km}^2$, as shown in Fig. \ref{2usertrajectory}. We set the flying altitude of the UAV as $H=100$~m, the reference channel power gain as $\beta_0=-30$~dB, the pathloss exponent as $\alpha=2$, the noise power as $\sigma^2_k=-50$~dBm$,\forall k\in{\mathcal K}$, the maximum UAV speed as $V_\text{max}=50$~m/s, the maximum transmit power of the UAV as $P=30~\text{dBm}$, the maximum transmit power of GUs as $Q_k=30~\text{dBm},~k\in{\mathcal K}$, and the minimum rate threshold of GUs as $\Gamma_k=2~\text{bps/Hz},~\forall k\in{\mathcal K}$. {We consider a simplified path loss model for the ground wireless channels as ${g}_{k}=\theta_0({\Theta}_{k}/d_0)^{-\epsilon},~\forall k\in{\mathcal K}$, where $\epsilon=3$ is the pathloss exponent, $\theta_0=-40$~dB corresponds to the pathloss at the reference distance of $d_0 = 1$ m, ${\Theta}_{k}$ denotes the distance from the GBS to its associated GU.} The mission period ${\mathcal T}$ is discretized into $N=200$ time slots.


{
Fig. \ref{2usertrajectory} shows the optimized UAV trajectory projected on the horizontal plane with $T = 150$ s. It is observed that for the proposed design, the UAV first reaches a point between GBSs 1 and 2, then flies towards GBS 3, and hovers there for around $92$ s. The UAV is observed to first connect with GBS 1 from 0 to $18.75$ s, then with both GBSs 1 and 2 from $18.75$ s to $24.75$ s, and finally with GBS 3, thus maximizing the UAV communication performance while protecting the GUs' communications.
By contrast, for the {\it{egoistic decoding scheme}}, the UAV is observed to fly close to GBS 1 and GBS 3, as it needs to connect with them in time interval $[0,27.75$ s$]$ and $(27.75$ s$,$~$150$ s$]$, respectively.
For the {\it{altruistic decoding scheme}}, the UAV is observed to fly among the three GBSs, as all the GBSs need to decode the messages of the UAV.
For the {\it{straight-fly scheme}}, the UAV is observed to connect with GBSs 1 and 3 during the whole period $\mathcal T$.
For the {\it{successive-hover-fly scheme}}, the UAV successively visits GBSs 2, 1, and 3, and hovers above GBS 3 for around $88.25$ s.}


{Fig. \ref{2userperformance} shows the average throughput of the UAV versus the mission duration $T$. It is observed that as $T$ becomes larger, the UAV's average throughput increases for all the five schemes except for the {\it{upper bound with $T\rightarrow\infty$}}, as the UAV can better exploit the mobility via hovering at desired locations for longer durations. It is also observed that the proposed design achieves the highest throughput among the five schemes over all regimes of $T$ and approaches the performance {\it{upper bound with $T\rightarrow\infty$}}, by jointly exploiting both trajectory design and adaptive IC over time.
The {\it{egoistic decoding scheme}} is observed to perform worse than our proposed design, as the UAV needs to keep a certain distance from the GBSs other than the specific associated GBSs to meet the minimum rate requirement of GUs.
The {\it{successive-hover-fly scheme}} and the {\it{straight-fly scheme}} are observed to perform even worse, due to the ignorance of the UAV trajectory design.
Furthermore, the {\it{altruistic decoding scheme}} is observed to perform worst, as the UAV's transmission rate needs to be sufficiently low so that all GBSs can decode the UAV's messages.}

\begin{figure}[t]
\centering
\includegraphics[width=6.5cm]{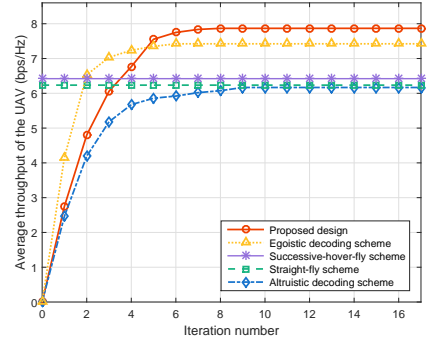}
\caption{{The convergence of the proposed design and benchmark schemes.}}\label{convergence}
\end{figure}

{
Fig. \ref{convergence} shows the convergence of the proposed design and the benchmark schemes. It is observed that the proposed design takes around $9$ iterations to convergence; while the {\it{egoistic decoding scheme}} and the {\it{altruistic decoding scheme}} respectively take $6$ and $10$ iterations to convergence.}

{\color{red}\begin{figure}[t]
\centering
\includegraphics[width=6.5cm]{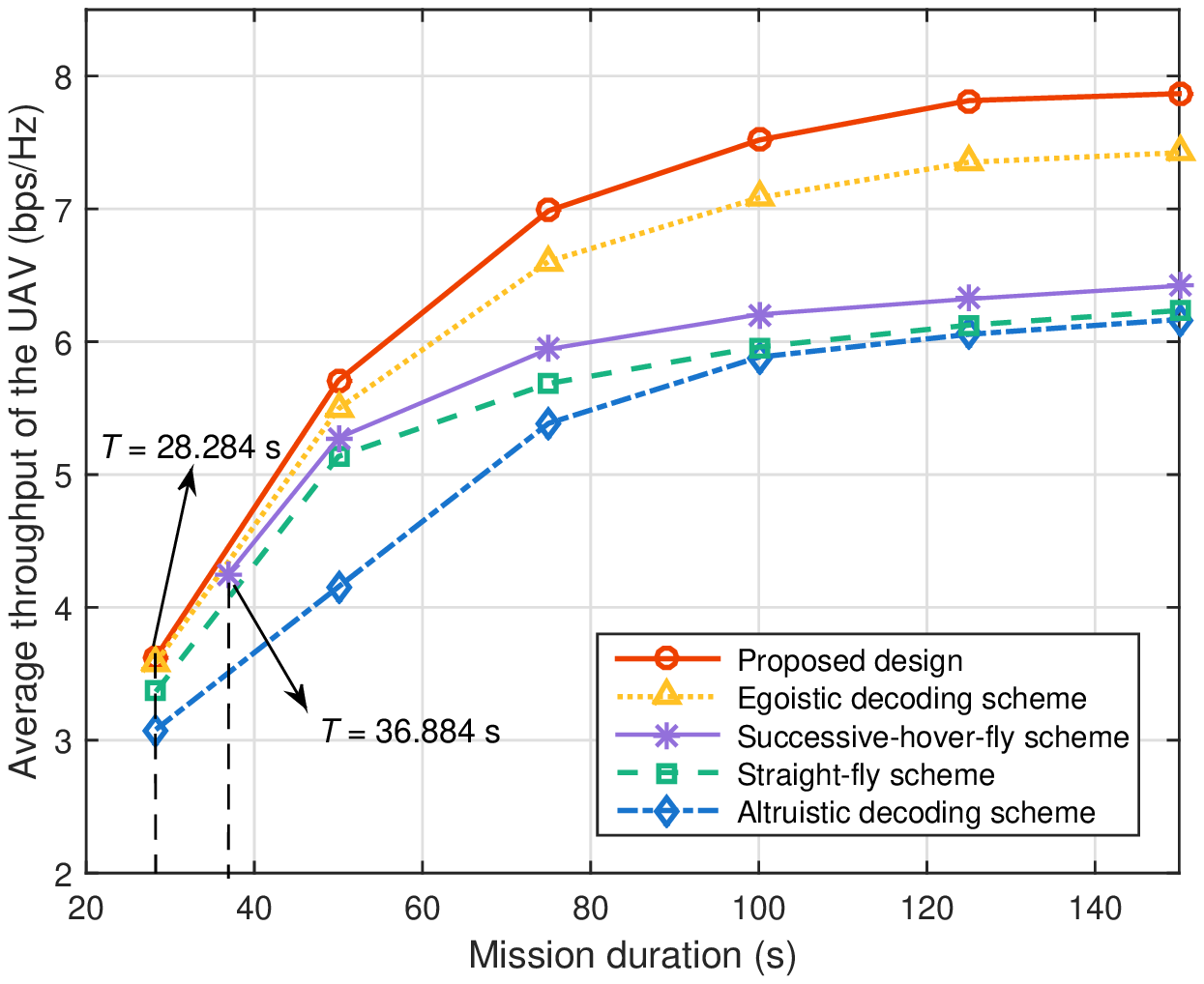}
\caption{{The performance versus different values of the UAV's mission duration $T$.}}\label{fea_T}
\includegraphics[width=6.5cm]{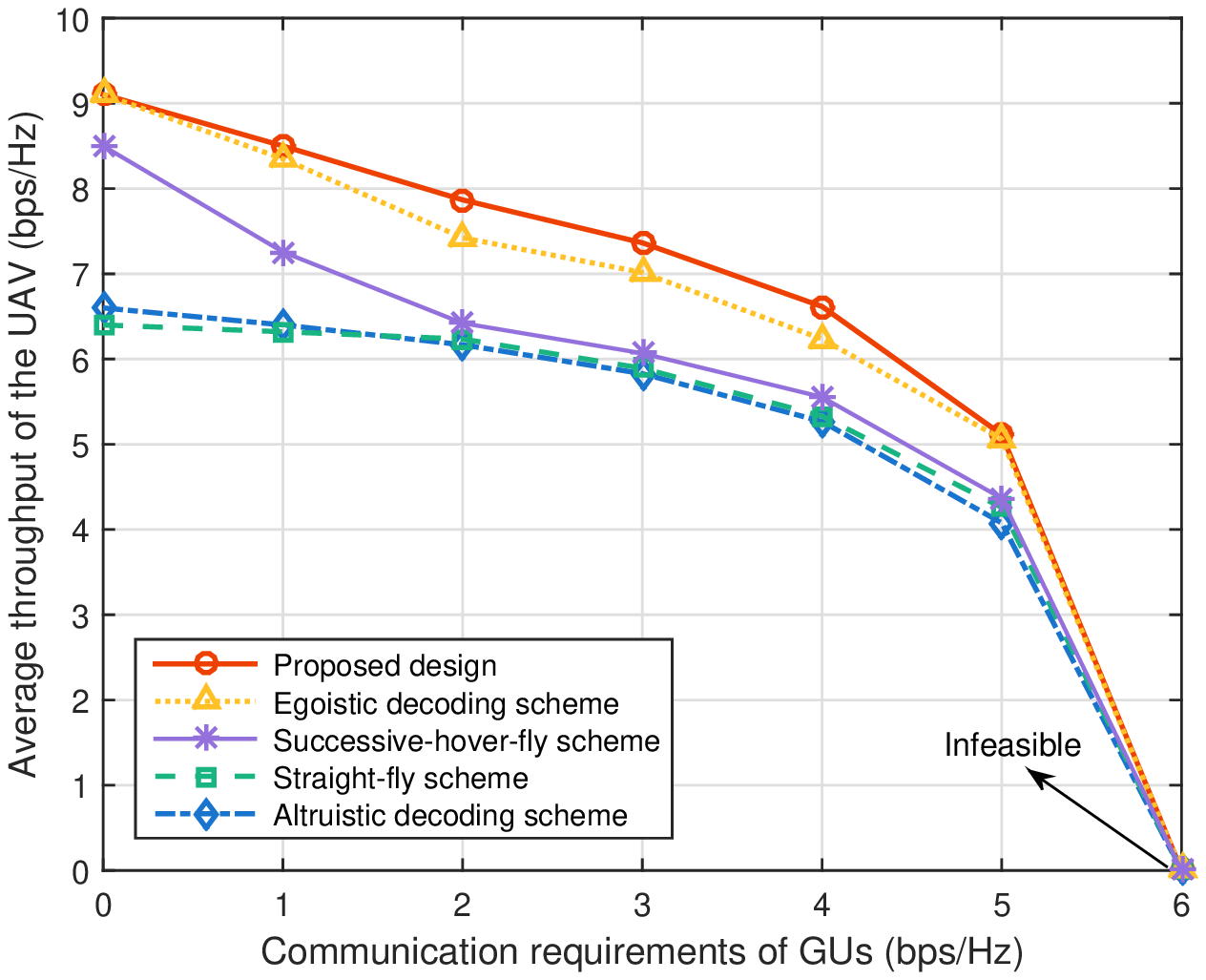}
\caption{{The performance versus different values of the communication requirements of GUs $\Gamma_k$.}}\label{feasibility_gamma}
\end{figure}}
{In order to demonstrate the feasibility of the considered problem under the above setup, we provide numerical results to respectively show the feasible ranges of the UAV's mission duration $T$ and the communication requirements of GUs $\Gamma_k$, as shown in Figs. \ref{fea_T} and \ref{feasibility_gamma}. Fig. \ref{fea_T} shows the performance versus different values of the UAV's mission duration $T$. It is observed that for the {\it{successive-hover-fly scheme}}, it takes at least $36.884$ s to visit all the three GBSs and hover above them. For the other four schemes, the minimum time requirement is $28.284$ s. As a result, under the considered setups, the feasible ranges of the UAV's mission duration $T$ for the {\it{successive-hover-fly scheme}} and the other four schemes are $[36.884~\text{s},T_\text{max})$ and $[28.284~\text{s},T_\text{max})$, respectively. Fig. \ref{feasibility_gamma} shows the performance versus different values of the communication requirement of GUs $\Gamma_k$. It is also observed that all the schemes are feasible when the communication requirement $\Gamma_k$ is no greater than $5$ bps/Hz. Accordingly, the feasible range of the communication requirements of GUs $\Gamma_k$ under the considered setups is $(0,5~\text{bps}]$.}

\begin{figure}[t]
\centering
\includegraphics[width=6.5cm]{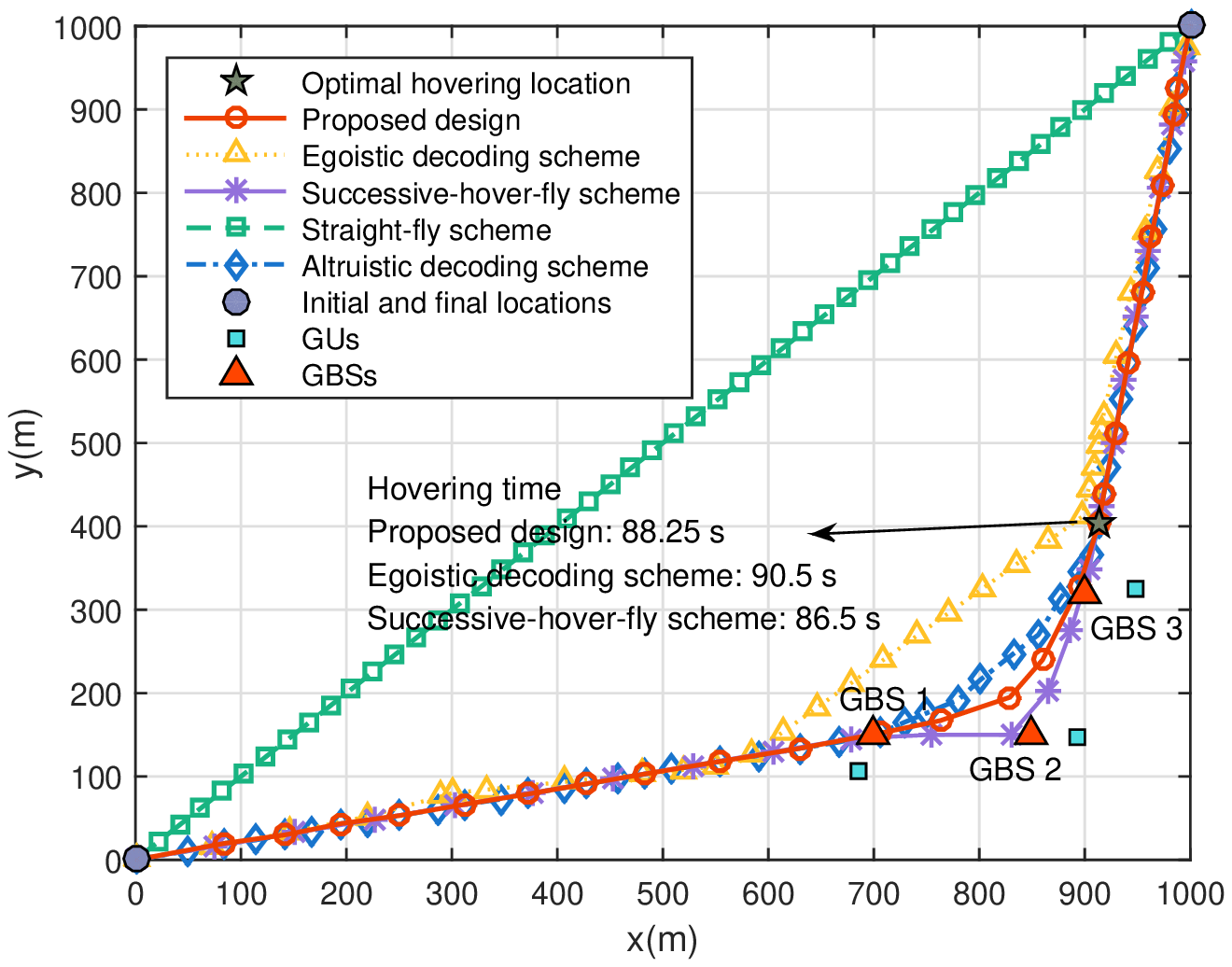}
\caption{{The optimized UAV trajectory projected on the horizontal plane with $T=150$ s under a different setup.}}\label{111}
\includegraphics[width=6.5cm]{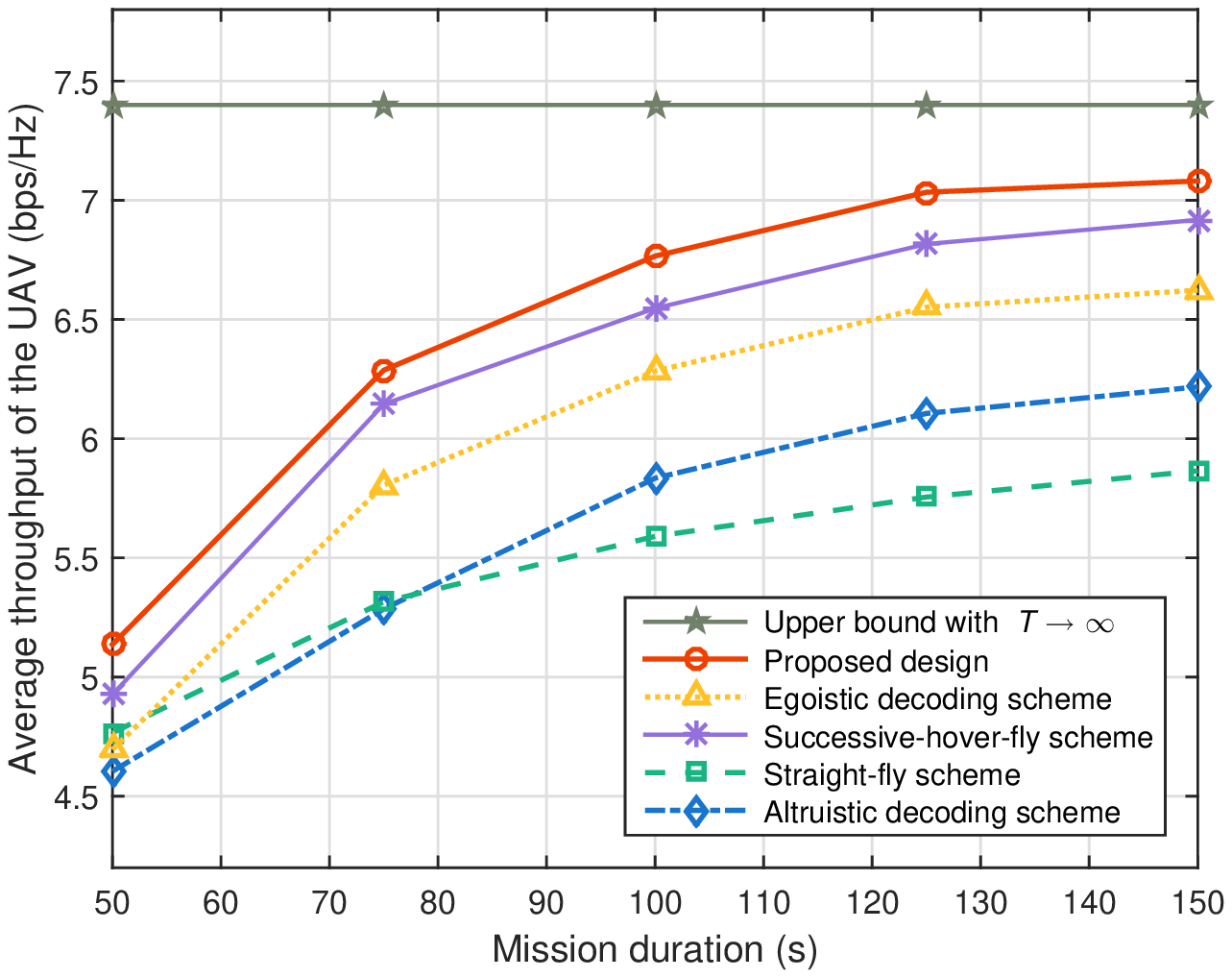}
\caption{{The average throughput of the UAV versus mission duration $T$ under a different setup.}}\label{222}
\end{figure}
{
Furthermore, we validate the performance of the proposed design under a different setup as shown in Figs. \ref{111} and \ref{222}.
Fig. \ref{111} shows the optimized UAV trajectory projected on the horizontal plane with $T = 150$ s based on a different setup. It is observed that the UAV first flies to visit GBSs 1, 2, and 3, then flies towards the final location for all the schemes except the {\it{straight-fly scheme}}, due to the fact that the three GBSs are distributed in a relatively close range. And the hovering locations for these schemes are the same as the optimal hovering locations corresponding to the case with $T\rightarrow\infty$.
Fig. \ref{222} shows the corresponding average throughput of the UAV versus the mission duration $T$. Similar observations can be found as in Fig. \ref{2userperformance} that as $T$ becomes larger, the UAV's average throughput increases for all the five schemes. It is also observed that the proposed design achieves the highest throughput among the five schemes over all regimes of $T$ and approaches the performance upper bound with $T\rightarrow\infty$.
In addition, in such a case where the GBSs are closely distributed, the performance gaps between the proposed scheme and the benchmark schemes are relatively reduced.}

\section{Conclusion}\label{sec:VI}
This letter studied an uplink spectrum sharing scenario for a cellular-connected UAV, in which a UAV user communicates with GBSs by sharing the spectrum with GUs. We proposed a new approach that jointly exploits the trajectory optimization and adaptive IC to maximize the data-rate throughput of the UAV, while protecting the communication data-rate of each GU. {How to extend the results to the cases with multiple UAVs, spectrum sharing of both uplink and downlink, more practical channel models, 3D UAV trajectory design, or online designs are interesting directions worth pursuing in future research.}

\appendices
{\section{Convergence Proof of the Proposed Algorithm} \label{Appendix}
To verify the convergence of the proposed alternating optimization based algorithm in Table I, in the following we show that in each iteration $i\ge1$, the obtained objective values of problem (P1) are monotonically non-decreasing.
For notational convenience, let $\hat{\mv{A}}^{(i)}=\{\{\hat{\tau_{k}}^{(i)}[n]\},\{\hat{q}^{(i)}_k[n]\},\{\hat{p}^{(i)}[n]\}\}$, $\hat{\mv{U}}^{(i)}=\{\hat{\mv{u}}^{(i)}[n]\}$, and $\hat{\mv{r}}^{(i)}=\{\hat{{r}}^{(i)}[n]\}$ denote the obtained resource allocation, UAV trajectory, and UAV's transmission rate at each outer iteration $i$ of alternating optimization.
Accordingly, we denote $\hat{R}^\text{avg}(\hat{\mv{A}}^{(i)},\hat{\mv U}^{(i)},\hat{\mv{r}}^{(i)})$ and $\hat{R}_\textrm{RA}^\text{avg}(\hat{\mv{A}}^{(i)},\hat{\mv U}^{(i)},\hat{\mv{r}}^{(i)})$ as the correspondingly achieved objective values of problem (P1) and problem (P1.1) based on $\hat{\mv{A}}^{(i)}$, $\hat{\mv{U}}^{(i)}$, and $\hat{\mv{r}}^{(i)}$, respectively.
Let ${\mv{U}}^{(j)}=\{{\mv{u}}^{(j)}[n]\}$ and ${\mv{r}}^{(j)}=\{{{r}}^{(j)}[n]\}$ denote the obtained UAV trajectory and UAV's transmission rate at each inner iteration $j$ of SCA and ${R}_\textrm{TO}^\text{avg}(\hat{\mv{A}}^{(i)},{{\mv U}}^{(j)},{\mv{r}}^{(j)})$ as the correspondingly achieved objective value of problem (P1.5) based on $\hat{\mv{A}}^{(i)}$, ${\mv{U}}^{(j)}$, and ${\mv{r}}^{(j)}$, respectively.

First, note that for any outer iteration $i\ge1$, the resource allocation $\hat{\mv{A}}^{(i)}$ and the UAV's transmission rate $\hat{\mv{r}}^{(i)}$ correspond to the optimal solution to problem (P1.1) under given UAV trajectory $\hat{\mv{U}}^{(i-1)}$. Therefore, we have
\begin{align}
\!\!\!\!\hat{R}^\text{avg}(\hat{\mv{A}}^{(i)},{\hat{\mv U}}^{(i-1)},\hat{\mv{r}}^{(i)})&=\hat{R}^\text{avg}_\text{RA}(\hat{\mv{A}}^{(i)},\hat{\mv U}^{(i-1)},\hat{\mv{r}}^{(i)})\notag\\
&\geq \hat{R}^\text{avg}_\text{RA}(\hat{\mv{A}}^{(i-1)},\hat{\mv U}^{(i-1)},\hat{\mv{r}}^{(i-1)})\notag\\
&=\hat{R}^\text{avg}(\hat{\mv{A}}^{(i-1)},\hat{\mv U}^{(i-1)},\hat{\mv{r}}^{(i-1)}).\label{AppA}
\end{align}
Next, we consider the update of the UAV trajectory $\hat{\mv U}^{(i)}$ and the UAV's transmission rate $\hat{\mv{r}}^{(i)}$ to problem (P1.4) under given resource allocation $\hat{\mv{A}}^{(i)}$, by solving its approximated problem (P1.5) via SCA. Consider the SCA initialized as ${\mv U}^{(0)}=\hat{\mv U}^{(i-1)}$. Then, for each inner iteration $j\ge1$, since the first-order Taylor expansions in (27) and (31) are tight at the given local UAV trajectory points ${\mv U}^{(j-1)}$, respectively, we have
\begin{align}
\!\!\!\!\hat{R}^\text{avg}(\hat{\mv A}^{(i)},{\mv U}^{(j-1)},\mv{r}^{(j-1)})\!=\!{R}^\text{avg}_\text{TO}(\hat{\mv A}^{(i)},{\mv U}^{(j-1)},\mv{r}^{(j-1)}).\label{AppD}
\end{align}
Furthermore, as the UAV trajectory ${\mv U}^{(j)}$ and the UAV's transmission rate ${\mv{r}}^{(j)}$ are the optimal solution to problem (P1.5) under given resource allocation $\hat{\mv{A}}^{(i)}$ and local UAV trajectory points ${\mv U}^{(j-1)}$, we have
\begin{align}
{R}^\text{avg}_\text{TO}(\hat{\mv A}^{(i)},{\mv U}^{(j)},\mv{r}^{(j)})\geq {R}^\text{avg}_\text{TO}(\hat{\mv A}^{(i)},{\mv U}^{(j-1)},\mv{r}^{(j-1)}). \label{AppE}
\end{align}
By combining \eqref{AppD} and \eqref{AppE}, it follows that
\begin{align}
\hat{R}^\text{avg}(\hat{\mv A}^{(i)},{\mv U}^{(j)},\mv{r}^{(j)})\geq\hat{R}^\text{avg}(\hat{\mv A}^{(i)},{\mv U}^{(j-1)},\mv{r}^{(j-1)}).
\label{AppB}
\end{align}
In other words, the obtained objective value of problem (P1.4) is monotonically non-decreasing for each inner iteration. As the objective value of problem (P1.4) is upper bounded, it is clear that the SCA-based update will lead to a converged solution to problem (P1.4). As such, we have
\begin{align}
\hat{R}^\text{avg}(\hat{\mv A}^{(i)},\hat{\mv U}^{(i)},\hat{\mv{r}}^{(i)})=\hat{R}^\text{avg}(\hat{\mv A}^{(i)},{\mv U}^{(j)},{\mv{r}}^{(j)}). \label{AppC}
\end{align}
By combining \eqref{AppA}, \eqref{AppB}, and \eqref{AppC}, it is clear that
\begin{align}
\hat{R}^\text{avg}(\hat{\mv A}^{(i)},\hat{\mv U}^{(i)},\hat{\mv{r}}^{(i)})\geq\hat{R}^\text{avg}(\hat{\mv{A}}^{(i-1)},\hat{\mv U}^{(i-1)},\hat{\mv{r}}^{(i-1)}),
\end{align}
which indicates that the objective value of problem (P1) is monotonically non-decreasing after each outer iteration of the alternating-optimization-based algorithm.
As a result, the convergence of the proposed algorithm in Table I is finally proved.}



\begin{thebibliography}{1}
\bibliographystyle{IEEEbib}
\bibitem{zeng1}
Y. Zeng, J. Lyu, and R. Zhang, ``Cellular-connected UAV: Potential, challenges, and promising technologies,'' {\it{IEEE Wireless Commun.}}, vol. 26, no. 1, pp. 120--127, Feb. 2019.
\bibitem{JSAC}
Q. Wu, J. Xu, Y. Zeng, D. W. K. Ng, N. Al-Dhahir, R. Schober, and A. L. Swindlehurst, ``A comprehensive overview on 5G-and-beyond networks with UAVs: From communications to sensing and intelligence,'' {\it{IEEE J. Sel. Areas Commun.}}, vol. 39, no. 10, pp. 2912--2945, Oct. 2021.
\bibitem{zeng2}
Y. Zeng, Q. Wu, and R. Zhang, ``Accessing from the sky: A tutorial on UAV communications for 5G and beyond,'' {\it{Proc. IEEE}}, vol. 107, no. 12, pp. 2327--2375, Dec. 2019.
\bibitem{R1}
3GPP-TR-36.777, ``Enhanced LTE support for aerial vehicles,'' 3GPP Tech. Rep., 2017. [Online]. Available: {\url{http://www.3gpp.org/dynareport/36777.htm}}

\bibitem{weidong3}
W. Mei and R. Zhang, ``Aerial-ground interference mitigation for cellular-connected UAV,'' {\it{IEEE Wireless Commun.}}, vol. 28, no. 1, pp. 167--173, Feb. 2021.
\bibitem{weidong2}
W. Mei, Q. Wu, and R. Zhang, ``Cellular-connected UAV: Uplink association, power control and interference coordination,'' {\it{IEEE Trans. Wireless Commun.}}, vol. 18, no. 11, pp. 5380--5393, Nov. 2019.
\bibitem{R3}
X. Cai, I. Z. Kov\'{a}cs, J. Wigard, R. Amorim, F. Tufvesson, and P. E. Mogensen, ``On the scheduling and power control for uplink cellular-connected UAV communications,'' 2021. [Online]. Available: {\url{https://arxiv.org/abs/2107.11738}}
\bibitem{guiguan}
X. Pang, G. Gui, N. Zhao, W. Zhang, Y. Chen, Z. Ding, and F. Adachi, ``Uplink precoding optimization for NOMA cellular-connected UAV networks,'' {\it{IEEE Trans. Commun.}}, vol. 68, no. 2, pp. 1271--1283, Feb. 2020.
\bibitem{liuliang}
L. Liu, S. Zhang, and R. Zhang, ``Multi-beam UAV communication in cellular uplink: Cooperative interference cancellation and sum-rate maximization,'' {\it{IEEE Trans. Wireless Commun.}}, vol. 18, no. 10, pp. 4679--4691, Oct. 2019.
\bibitem{weidong}
W. Mei and R. Zhang, ``Uplink cooperative NOMA for cellular-connected UAV,'' {\it{IEEE J. Sel. Top. Signal Process.}}, vol. 13, no. 3, pp. 644--656, Jun. 2019.
\bibitem{weidong4}
W. Mei and R. Zhang, ``Uplink cooperative interference cancellation for cellular-connected UAV: A quantize-and-forward approach,'' {\it{IEEE Wireless Commun. Lett.}}, vol. 9, no. 9, pp. 1567--1571, Sept. 2020.
{\bibitem{anti1}
B. Duo, Q. Wu, X. Yuan, and R. Zhang, ``Anti-jamming 3D trajectory design for UAV-enabled wireless sensor networks under probabilistic LoS channel,'' {\it{IEEE Trans. Veh. Technol.}}, vol. 69, no. 12, pp. 16288--16293, Dec. 2020.}
{
\bibitem{anti2}
H. Wang, J. Chen, G. Ding, and J. Sun, ``Trajectory planning in UAV communication with jamming,'' in {\it{Proc. WCSP}}, 2018, pp. 1--6.}
\bibitem{yuwei}
Y. Huang, W. Mei, J. Xu, L. Qiu, and R. Zhang, ``Cognitive UAV communication via joint maneuver and power control,'' {\it{IEEE Trans. Commun.}}, vol. 67, no. 11, pp. 7872--7888, Nov. 2019.
\bibitem{RL}
U. Challita, W. Saad, and C. Bettstetter, ``Interference management for cellular-connected UAVs: A deep reinforcement learning approach,'' {\it{IEEE Trans. Wireless Commun.}}, vol. 18, no. 4, pp. 2125--2140, Apr. 2019.
{
\bibitem{channel}
D. W. Matolak and R. Sun, ``Air-ground channel characterization for unmanned aircraft systems-Part III: The suburban and near-urban environments,'' {\it{IEEE Trans. Veh. Technol.}}, vol. 66, no. 8, pp. 6607--6618, Aug. 2017.}

{
\bibitem{online}
A. Khalili, E. M. Monfared, S. Zargari, M. R. Javan, N. Mokari, and E. A. Jorswieck, ``Resource management for transmit power minimization in UAV-assisted RIS HetNets supported by dual connectivity,'' to appear in {\it{IEEE Trans. Wireless Commun.}}, 2022.}

{
\bibitem{online1}
C. You and R. Zhang, ``Hybrid offline-online design for UAV-enabled data harvesting in probabilistic LoS channels,'' {\it{IEEE Trans. Wireless Commun.}}, vol. 19, no. 6, pp. 3753--3768, Jun. 2020.}
{\bibitem{online2}
Y. Sun, D. Xu, D. W. K. Ng, L. Dai, and R. Schober, ``Optimal 3D-trajectory design and resource allocation for solar-powered UAV communication systems,'' {\it{IEEE Trans. Commun.}}, vol. 67, no. 6, pp. 4281--4298, Jun. 2019.}
{\bibitem{online3}
C. Zhan and Y. Zeng, ``Energy-efficient data uploading for cellular-connected UAV systems,'' {\it{IEEE Trans. Wireless Commun.}}, vol. 19, no. 11, pp. 7279--7292, Nov. 2020.}

{
\bibitem{altitude2}
B. Hu, L. Wang, S. Chen, J. Cui, and L. Chen, ``An uplink throughput optimization scheme for UAV-enabled urban emergency communications,'' to appear in {\it{IEEE Internet Things J.}}, 2022.}
{
\bibitem{TSP}
L. Xie, J. Xu, and R. Zhang, ``Throughput maximization for UAV-enabled wireless powered communication networks,'' {\it{IEEE Internet Things J.}}, vol. 6, no. 2, pp. 1690--1703, Apr. 2019.}

\bibitem{boyd}
S. Boyd and L. Vandenberghe, {\em Convex optimization}, Cambridge, U.K.: Cambridge Univ. Press, Mar. 2004.
\bibitem{CVX}
M. Grant and S. Boyd, {\it{CVX: MATLAB Software for Disciplined Convex Programming}}, 2016. [Online] Available: {\url{https://cvxr.com/cvx}}


\end{thebibliography}
\end{document}